\documentclass[%
reprint, twocolumn, showpacs, longbibliography,
superscriptaddress,
amsmath, amssymb,
aps,
floatfix,
]{revtex4-2}

\usepackage{graphicx}
\usepackage{dcolumn}
\usepackage{bm}
\usepackage{color}
\usepackage{txfonts}
\usepackage{microtype}
\usepackage{siunitx}
\usepackage{url}
\usepackage{epstopdf}
\usepackage{indentfirst}
\usepackage[colorlinks, linkcolor=black, anchorcolor=black, citecolor=black, 
            filecolor=black, urlcolor=blue, bookmarksopen=true]{hyperref}
\usepackage{textcomp}
\usepackage{mathrsfs}
\usepackage{braket}
\usepackage[caption=false, font=normalsize, labelfont=sf, textfont=sf]{subfig}

\begin{document}

\title{Efficient Generation of Neutrons Based on Ultrashort Laser-driven Direct Acceleration in Microwire-Array Targets} 

\author{K. Y. Feng}
\thanks{These authors contributed equally to this work.}
\affiliation{College of Science, National University of Defense Technology, Changsha 410073, China}

\author{D. B. Zou}
\thanks{These authors contributed equally to this work.}
\affiliation{College of Science, National University of Defense Technology, Changsha 410073, China}

\author{B. Cui}
\affiliation{Science and Technology on Plasma Physics Laboratory, Laser Fusion Research Center, China Academy of Engineering Physics, Mianyang 621900, China}

\author{S. K. He}
\affiliation{Science and Technology on Plasma Physics Laboratory, Laser Fusion Research Center, China Academy of Engineering Physics, Mianyang 621900, China}

\author{Y. Z. Dai}
\affiliation{State Key Laboratory of High Field Laser Physics, Shanghai Institute of Optics and Fine Mechanics, Chinese Academy of Sciences, Shanghai 201800, China}

\author{W. Qi}
\affiliation{Science and Technology on Plasma Physics Laboratory, Laser Fusion Research Center, China Academy of Engineering Physics, Mianyang 621900, China}

\author{J. L. Luo}
\affiliation{College of Science, National University of Defense Technology, Changsha 410073, China}

\author{J. Feng}
\affiliation{College of Science, National University of Defense Technology, Changsha 410073, China}

\author{X. Y. Li}
\affiliation{College of Science, National University of Defense Technology, Changsha 410073, China}

\author{Z. H. Chen}
\affiliation{College of Science, National University of Defense Technology, Changsha 410073, China}

\author{L. X. Hu}
\affiliation{College of Science, National University of Defense Technology, Changsha 410073, China}

\author{C. Y. Qin}
\affiliation{State Key Laboratory of High Field Laser Physics, Shanghai Institute of Optics and Fine Mechanics, Chinese Academy of Sciences, Shanghai 201800, China}

\author{G. B. Zhang}
\affiliation{College of Science, National University of Defense Technology, Changsha 410073, China}

\author{H. Zhang}
\affiliation{State Key Laboratory of High Field Laser Physics, Shanghai Institute of Optics and Fine Mechanics, Chinese Academy of Sciences, Shanghai 201800, China}

\author{Z. G. Deng}
\affiliation{Science and Technology on Plasma Physics Laboratory, Laser Fusion Research Center, China Academy of Engineering Physics, Mianyang 621900, China}

\author{X. H. Yang}
\affiliation{College of Science, National University of Defense Technology, Changsha 410073, China}

\author{F. Q. Shao}
\affiliation{College of Science, National University of Defense Technology, Changsha 410073, China}

\author{L. L. Ji}
\email{jill@siom.ac.cn}
\affiliation{State Key Laboratory of High Field Laser Physics, Shanghai Institute of Optics and Fine Mechanics, Chinese Academy of Sciences, Shanghai 201800, China}

\author{W. M. Zhou}
\email{zhouwm@caep.cn}
\affiliation{Science and Technology on Plasma Physics Laboratory, Laser Fusion Research Center, China Academy of Engineering Physics, Mianyang 621900, China}

\author{T. P. Yu}
\email{tongpu@nudt.edu.cn}
\affiliation{College of Science, National University of Defense Technology, Changsha 410073, China}

\date{\today}

\begin{abstract}

We report on an experimental demonstration of efficient neutron generation based on direct laser acceleration in microwire-array targets irradiated by ultrashort (tens of femtoseconds) laser pulses. The optimal array period was identified, at which the maximum proton energy and the number of protons with energies exceeding 1~MeV were significantly increased. Using a 1~PW, $\sim$25~fs laser at a moderate intensity of $\sim10^{20}$~W/cm$^{2}$, a high neutron yield of up to (8.33~$\pm$~0.84)~$\times$~$10^{6}$~n/sr/J was detected from the LiD converter via $^7\text{Li}(p,n)$ and $\text{D}(p,n+p)$ nuclear reactions. Self-consistent integrated simulations reproduced the experimental results and predicted that with a Be converter, a forward pulsed neutron source with an unprecedented yield per joule of 3.67~$\times$~$10^{7}~$n/sr/J can be obtained under identical laser conditions. This type of neutron source is favorable for applications that require a high repetition rate utilizing compact and economical laser systems.  

\end{abstract}

\maketitle


Short duration, high flux neutron sources are particularly favored in applications such as neutron resonance imaging ~\cite{vartsky2010novel}, neutron physics of fusion materials ~\cite{perkins2000investigation}, and researches on fast neutron capture ~\cite{wallerstein1997synthesis}. Until now, neutron sources with such characteristics primarily relied on accelerator-based facilities. The peak neutron flux achieved in these facilities has reached 10$^{17}$~n/cm$^{2}$/s ~\cite{taylor2007route,vogel2012brief}, and the duration of neutron pulse is as short as the nanosecond scale ~\cite{altstadt2007photo}. However, it remains a challenging endeavor to produce a pulsed neutron source with higher flux and shorter duration due to limitations in energy consumption, technology and economic costs, etc.

For years now, neutron sources based on laser accelerator have attracted significant attention ~\cite{alvarez2014laser,yogo2023advances,canova2025high,gutberlet2026recent}, owing to the unique features such as an ultrashort duration ($<$1~ns) ~\cite{chen2019extreme,zweiback2000characterization,pomerantz2014ultrashort,higginson2015temporal,qi2019enhanced,yogo2023laser}, micro source size ($\sim$mm) ~\cite{guler2016neutron,jiang2021microstructure,jiao2023high}, and ultrahigh peak flux ($>$$10^{18}$~n/cm$^{2}$/s) ~\cite{chen2019extreme,treffert2021towards,jiao2023high}. These features enable it to serve as a powerful supplement to traditional neutron sources, and are expected to further expand the application scenarios of neutrons. In general, hydrogen isotopes such as protons and deuterium ions are preferred as the charged particle types for laser acceleration due to their high reaction cross sections. When they collide with each other or are directed onto the converters placed downstream, neutrons will be released through nuclear reactions. In the scenario where the laser-irradiated target is separated from the converter ~\cite{lancaster2004characterization,16}, the resulting neutron source has a relatively high yield ~\cite{jiao2023high,feng2020high,alejo2017high} and even exhibits excellent forward orientation ~\cite{alejo2017high,kar2016beamed,davis2010neutron,liu2022laser,kleinschmidt2018intense,roth2013bright}. Neutron yields exceeding $10^{10}$~n/sr have been reported in the experiments ~\cite{yogo2023laser,kleinschmidt2018intense,roth2013bright,maksimchuk2013dominant,yao2023high}, and the highest neutron yield per joule has reached about 6.25~$\times$~$10^{7}~$n/sr/J ~\cite{kleinschmidt2018intense,roth2013bright}. Unfortunately, achieving high neutron yield usually requires long-pulse lasers with hundreds of joules of energy ~\cite{yogo2023laser,feng2020high,yao2023high,bo2021experimental,favalli2025demonstration,zimmer2022demonstration,maksimchuk2013dominant}, and such laser facilities can operate only at a low repetition rate. Ultrashort (tens of femtoseconds) ultraintense laser pulses are advantageous in producing neutrons at a high repetition rate, and their laser systems are compact and economical. However, the neutron yield per joule obtained is unsatisfactory because laser-ion acceleration is not as efficient as high-energy long-pulse lasers~\cite{lelievre2024high}. So far, the highest neutron yield per joule measured in the experiments using moderate-intensity ($I\lesssim10^{20}\ \rm W/cm^{2}$) laser pulses of sub-100~fs duration is still no more than 3.08~$\times$~$10^{6}$~n/sr/J despite a great deal of effort has been devoted to improving the neutron production ~\cite{treffert2021towards,norreys1998neutron,lelievre2024high,disdier1999fast,youssef2005broad,hah2018characterization,kong2022high,knight2024detailed,wang2025high,stuhl2025continuous,curtis2018micro,jiang2020energetic,deng2022pulsed,miran2023laser,osvay2024fast,osvay2024towards,lei2024compact}.

\begin{figure*}
    \centering
    \includegraphics[width=0.9\hsize]{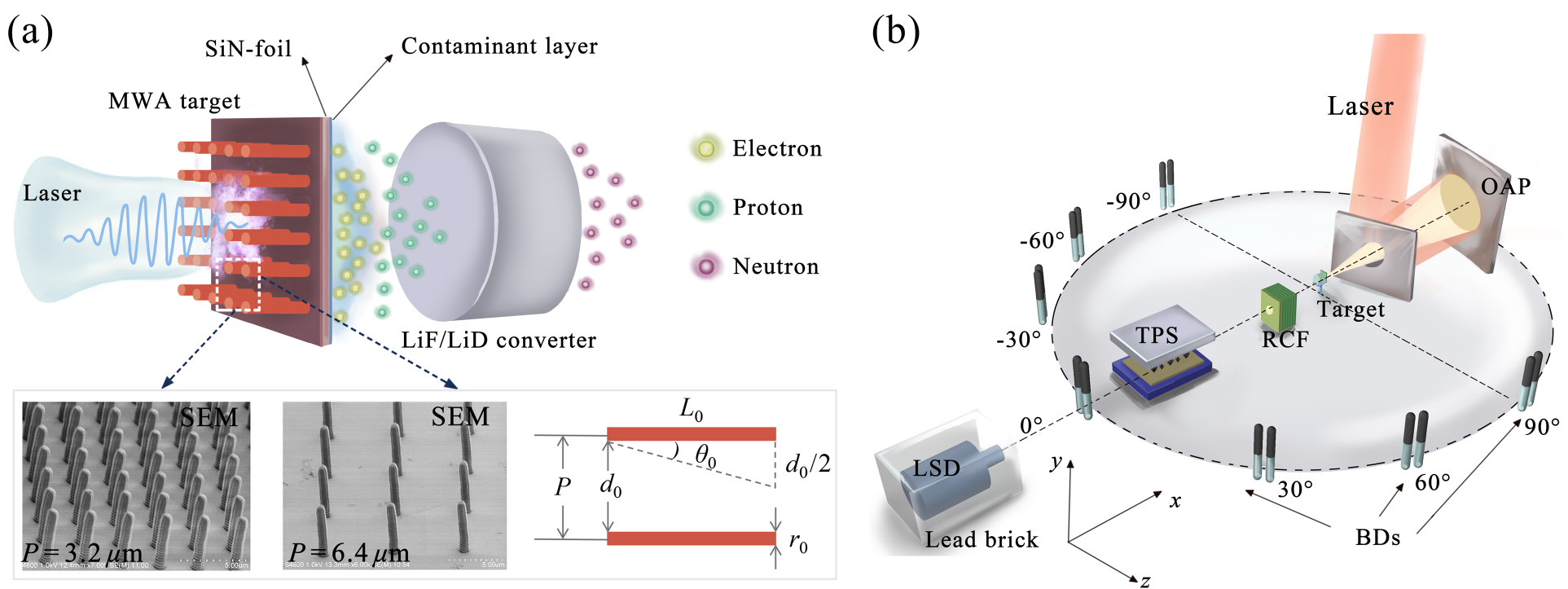} 
    \caption{(a) Schematic drawing of neutron generation based on direct laser acceleration. 3D-printed microwire arrays (MWAs) are attached to a SiN flat foil, and the LiF or LiD converters are placed 1 cm away. Inset shows the SEM images of microwire arrays with $P = 3.2 \ \mu\rm m$ and $6.4 \ \mu\rm m$ as well as the relevant parameters of the MWA size. (b) Diagnostics setup. Spatial-intensity profile and spectrum of protons were detected using stacked dosimetry films (RCF) and Thomson parabola spectrometer (TPS) at the target normal direction. Bubble detectors (BDs) were arranged along the chamber surface from -90° to 90° at 30° intervals to measure the neutron yield in each angular direction. To ensure accuracy, two BDs were deployed at each angular position, all positioned 62 cm from the target. A liquid scintillation detector (LSD) was aligned along the 0° direction to record the neutron energy spectrum. To effectively mitigate x-ray contamination of the neutron signal, the LSD was shielded by 20 cm thick lead bricks.}
    \label{fig1}
\end{figure*}

In this Letter, we report on the experimental generation of neutrons based on direct laser acceleration (DLA) in ultrashort laser interacting with microwire-array (MWA) targets. The MWA target is composed of highly ordered microwires and an attached substrate. As the laser enters the MWA, the DLA occurs inside the channels constituted by the neighbouring microwires. The overdense electron bunches pulled out of the wires by the laser electric field can gain forward momenta via the Lorentz force ~\cite{pukhov1999particle,kluge2012high,jiang2016microengineering}. For appropriate wire lengths and array periods, these electrons can be accelerated to a superponderomotive temperature ~\cite{jiang2016microengineering,qin2022high}. When they penetrate through the substrate, an exceptionally intense and broad sheath electric field will be induced behind the substrate. A significant increase in proton and ion energies is achieved ~\cite{kluge2012high,qin2022high}, compared to the typical target normal sheath acceleration (TNSA) using conventional plane target ~\cite{hatchett2000electron,snavely2000intense,wilks2001energetic}. High neutron yield per joule is thus expected when a suitable converter is subsequently employed. To verify the applicability of DLA for assisting efficient neutron production, we carry out the proof-of-principle experiment by using 3D-printed MWA targets. A strong dependence of neutron yield on the array period is observed. Under the optimal period condition, neutron emission as high as (8.33~$\pm$~0.84)~$\times$~$10^{6}$~n/sr/J is measured from the LiD converter with laser pulses of power 1~PW and duration $\sim$25~fs, setting a new record of neutron yield for fs-class lasers at moderate intensities. Self-consistent integrated simulations incorporating the radiation hydrodynamic (RHD), particle-in-cell (PIC) and Monte Carlo (MC) methods have reproduced the experimental results and predicted that by using a Be converter, a pulsed neutron source with yield per joule of 3.66~$\times$~$10^{7}$~n/sr/J can be obtained under the same laser parameters. The breakthrough of neutron yield in excess of $10^{7}$~n/sr/J at a high repetition rate would be able to trigger significant advances in the applications related to high-flux short-pulse neutron sources, e.g., neutron resonance radiography~\cite{Williams2020fast,simon2025high} and fusion materials analysis ~\cite{perkins2000the}.  

The experiments were carried out on the SILEX-II laser facility ~\cite{hong2021commissioning}, which has a central wavelength of $\lambda_0=800$~nm and duration $\sim$25~fs, and delivers an energy $\mathscr{E}_{L}\approx 30\ \rm J$, with over 50\% of the energy concentrated within $10\ \mu\mathrm{m}$. The peak power of the laser reached about 1 PW with an intensity of $I\sim 10^{20}\ \rm W/cm^{2}$. The laser contrast ratio $(CR)$ was better than $10^{10}$ at 20 ps ahead of the main pulse. The schematic view of our experimental setup is depicted in Fig. 1(a). The laser beam was focused onto the front surface of the substrate with an incident angle of $5^{\circ}$. The microwire array consists of a polymer material of mass density 1.17 g/cm$^{3}$, and is 3D-printed onto a SiN flat substrate of thickness 200 nm ~\cite{qin2022high}. The length and diameter of microwires were fixed at  $L_0 = 6 \ \mu\rm m$ and  $r_0 = 1 \ \mu\rm m$, while the array period (i.e., transverse interval between the axes of two adjacent microwires) was varied at $P = 3.2 \ \mu\rm m$, $4.8 \ \mu\rm m$, $6.4 \ \mu\rm m$ and $7.2 \ \mu\rm m$. The structural stability of microwires were guaranteed through ultraviolet treatment ~\cite{stuhl2025continuous}. As representatives, the scanning electron microscopy (SEM) images of the orderly arranged microwires with  $P = 3.2 \ \mu\rm m$ and $6.4 \ \mu\rm m$ are presented in the inset of Fig.~\ref{fig1}(a). The LiF and LiD blocks of thickness $D$ = 1 cm are used as the converters, which are 1 cm away from the CH contaminant layer adhering to the rear surface of MWA target. The diagnostics setup is illustrated in Fig.~\ref{fig1}(b) and the details are described in the caption. In addition, the experimental shots for plane target combined with a converter (without microwire-array, w/o) were also performed for comparison.

\begin{figure*}
		\centering
		\includegraphics[width=0.92\hsize]{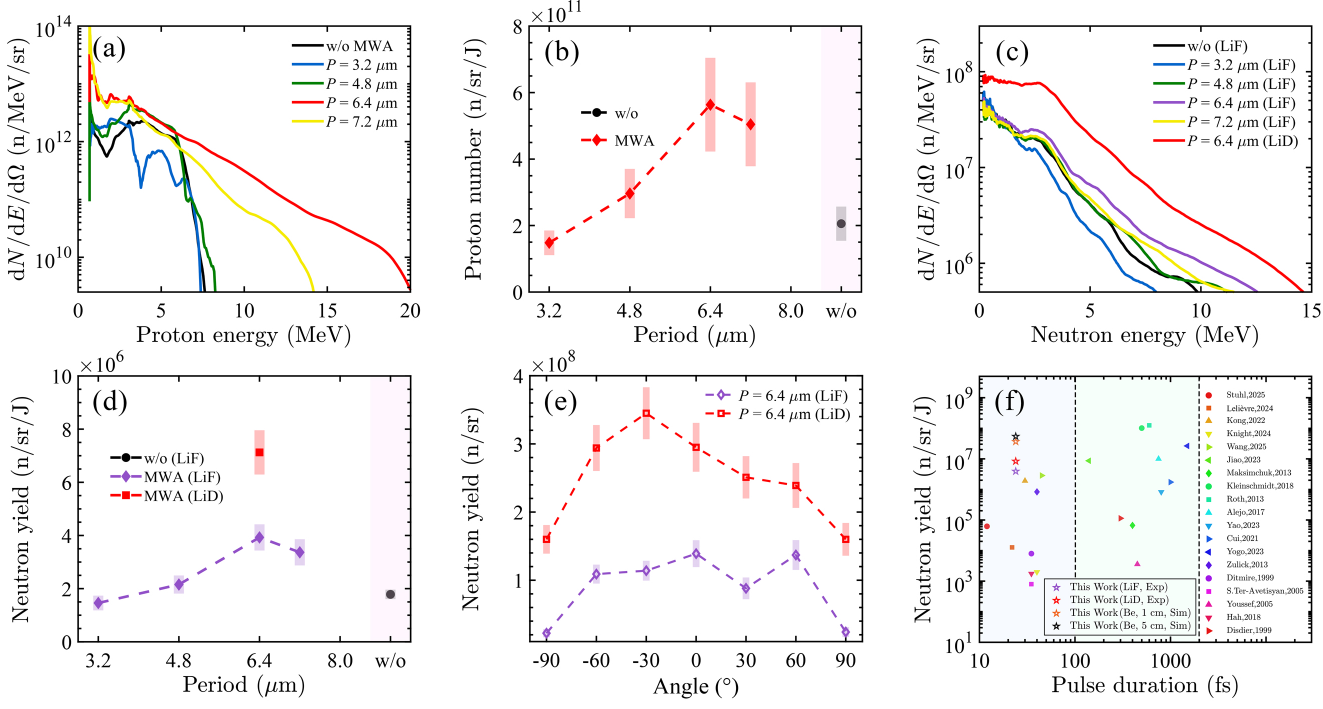} 
		\caption{Experimental results. (a), (b) Proton energy spectra measured with TPS and the total number of protons per joule of laser energy with different array period $P$. (c), (d) Neutron energy spectra and yield per joule obtained from liquid scintillation detector(LSD) and bubble detectors(BDs) (at the $0^\circ$ direction) at different $P$. The results for the case of plane target (without microwire-array, w/o) are also given in (a)-(d) for comparison. (e) Angular distribution of neutrons with MWA targets of $P = 6.4 \ \mu\rm m$ combined with the LiF and LiD converters. (f) Experimental neutron yield per joule (n/sr/J) based on moderate intensity laser-driven proton and ion sources as a function of laser pulse duration. Here, only the best results obtained from each laser facility are shown. The solid and hollow star-shaped symbols represent our experimental (Exp) and simulation (Sim) results, respectively.}
		\label{fig2}
	\end{figure*}

Figures~\ref{fig2}(a) and \ref{fig2}(b) plot the proton energy spectrum from TPS and the total number of protons per joule calculated through spectral integration in the MWA targets of different $P$ and plane target, respectively. Considering the proton energy thresholds of $^7\text{Li}(p,n)$ nuclear reaction, only the number of the protons with energies greater than 1 MeV are counted. We see that both the proton cutoff energy and number are highest at $P = 6.4 \ \mu\rm m$, which are about 3 times improved compared to the plane target. The neutron energy spectrum and yield detected from LSD and BDs at the 0º direction are shown in Figs.~\ref{fig2}(c) and \ref{fig2}(d). For the LiF converter, the neutron cutoff energy $\mathscr{E}_{n,\rm cutoff}$ and neutron yield per joule $Y_n/ \mathscr{E}_L$ released through $^7\text{Li}(p,n)$ reaction also reach their maximum values 12.5 MeV and (3.93~$\pm$~0.49)~$\times$~$10^{6}$~n/sr/J at $P = 6.4 \ \mu\rm m$. This yield is more than twice as high as (1.78~$\pm$~0.18)~$\times$~$10^{6}$~n/sr/J from the plane target. When a LiD converter is used, they are further increased since the $\text{D}(p,n+p)$ reaction will also be triggered, which remains a relatively high cross section within a broader proton energy range. At the optimal period of $P^{\rm Exp}_{\rm opt}= 6.4\ \mu \rm m$, $\mathscr{E}_{n,\rm cutoff}$ is approximately 15 MeV and $Y_n/ \mathscr{E}_L$ can reach $(7.12 \pm 0.84) \times 10^6$ n/sr/J. Furthermore, we noticed that more neutrons from the LiD converter are forward emitted, in contrast to the reference LiF converter [see Fig.~\ref{fig2}(e)]. This is due to the fact that the $\text{D}(p,n+p)$ reaction is dominated by direct nuclear reaction and the angular distribution of the neutrons produced is highly anisotropic, while the $^7\text{Li}(p,n)$ reaction belongs to compound nuclear reaction with the neutrons emitted symmetrically over the entire 4$\pi$ solid angle ~\cite{bohr1936neutron}. The highest neutron yield for the LiD converter is as high as $(3.43 \pm 0.35) \times 10^8$ n/sr [corresponding to a neutron yield per joule of $(8.33 \pm 0.84) \times 10^6$ n/sr/J], which is over one order of magnitude higher than the experimental record $2 \times 10^7$ n/sr using ultrashort laser pulses at moderate intensities~\cite{kong2022high}. Figure~\ref{fig2}(f) compiles the benchmarks of neutron yield based on moderate intensity laser-driven proton and ion beams in the experiments ~\cite{ditmire1999nuclear,zulick2013energetic,ter2005fusion,yogo2023laser,jiao2023high,alejo2017high,kleinschmidt2018intense,roth2013bright,maksimchuk2013dominant,yao2023high,bo2021experimental,disdier1999fast,youssef2005broad,hah2018characterization,kong2022high,knight2024detailed,lelievre2024high,wang2025high,stuhl2025continuous}, including from kHz high-repetition-rate femtosecond laser systems to kJ-class picosecond laser facilities. The observed neutron yield per joule $(8.33 \pm 0.84) \times 10^6$ n/sr/J in the current experiment is the highest using moderate intensity lasers with duration less than 100 fs.

To provide a self-consistent interpretation of the experimental observations, comprehensive simulations with the RHD code Flash ~\cite{fryxell2000flash}, PIC code Epoch ~\cite{arber2015contemporary} and MC code Geant4 ~\cite{shikhaliev2005quantum} have been performed to investigate the processes of preplasma formation, laser proton acceleration and neutron production, respectively. The preplasma profile obtained from RHD simulations is first applied to the PIC simulation as the initial ($\Delta t = 0$) density distribution of MWA target. All proton information from PIC simulations is then imported into MC simulations of nuclear reactions as the input. The simulations accounted for the cases of the MWA targets with different $P$ as well as the plane target (w/o). More details about the comprehensive simulations can be seen in Appendix A.

Figure~\ref{fig3} represents the distribution of the MWA density at $t=120$ ps from RHD simulations and the electron density at $\Delta t = 60T_0$ from PIC simulations with $P = 3.2 \ \mu\rm m$, $6.4\ \mu \rm m$ and $9.6\ \mu \rm m$, respectively. One can see that the microwires expand outward and undergo slight deformation due to the continuous ablation by laser prepulse. The expanded preplasmas gradually filled the channels between the two adjacent microwires. At $P=6.4\ \mu \rm m$ and $9.6\ \mu \rm m$, the regions of low-density plasma or vacuum still exist inside the channels, whereas the entire space covered by the laser focal spot is almost completely occupied by preplasmas as $P$ reduces to $3.2\ \mu \rm m$. When $P \geq 6.4\ \mu \rm m$, overdense electron bunches are pulled out by the transverse electric field force $e\vec{E_y}$ of the laser pulse as the laser propagates through the channels. Then they move forward under the action of the Lorentz force  ($e \vec{v} \times \vec{B}$) and further gain energies from the laser ~\cite{pukhov1999particle,kluge2012high,jiang2016microengineering,qin2022high}. It is noteworthy that fewer electrons are extracted from the sides ($y = \pm P$) due to the weaker laser field there compared to the axial region. This decline becomes more significant as $P$ increases. In contrast, for narrow channel entrances ($P \leq3.2\ \mu \rm m$), more laser energy will be reflected, leading to a considerable decrease of the electron number behind the substrate.

\begin{figure}
		\centering
		\includegraphics[width=1\hsize]{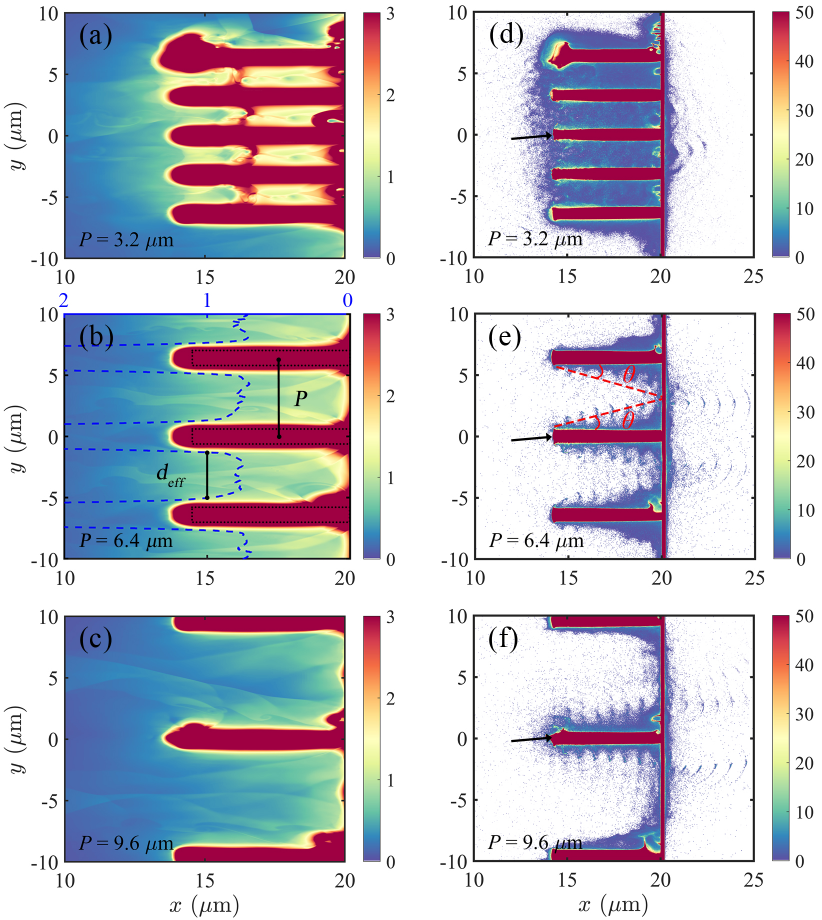} 
		\caption{Distributions of (a)-(c) the MWA density $t=120$ ps from RHD simulations, and (d)-(f) the electron density at $\Delta t=60 T_0$ from PIC simulations with $P = 3.2 \ \mu\rm m$, $6.4\ \mu \rm m$ and $9.6\ \mu \rm m$, respectively. All densities are normalized by the critical plasma density $n_c=m_e\omega_0^2/4\pi e^2$, where $\omega_0=2\pi/T_0$ is the central frequency and period of the laser, $T_0=\lambda_0/c$ is the corresponding laser period, $c$ is the speed of light in vacuum, and $-e$ and $m_e$ are the charge and rest mass of electron, respectively. In (b), $d_{\rm eff}$ denotes the effective wire distance evaluated at the critical density positions. The black dashed curves outline the initial wire positions, while the blue dashed curves represent the electron density profiles. The arrows in (d)–(f) indicate the incident direction of the laser, and $\theta$ in (e) represents the focusing angle of the electron bunches.}
		\label{fig3}
	\end{figure}
    
\begin{figure*}
		\centering
		\includegraphics[width=.9\hsize]{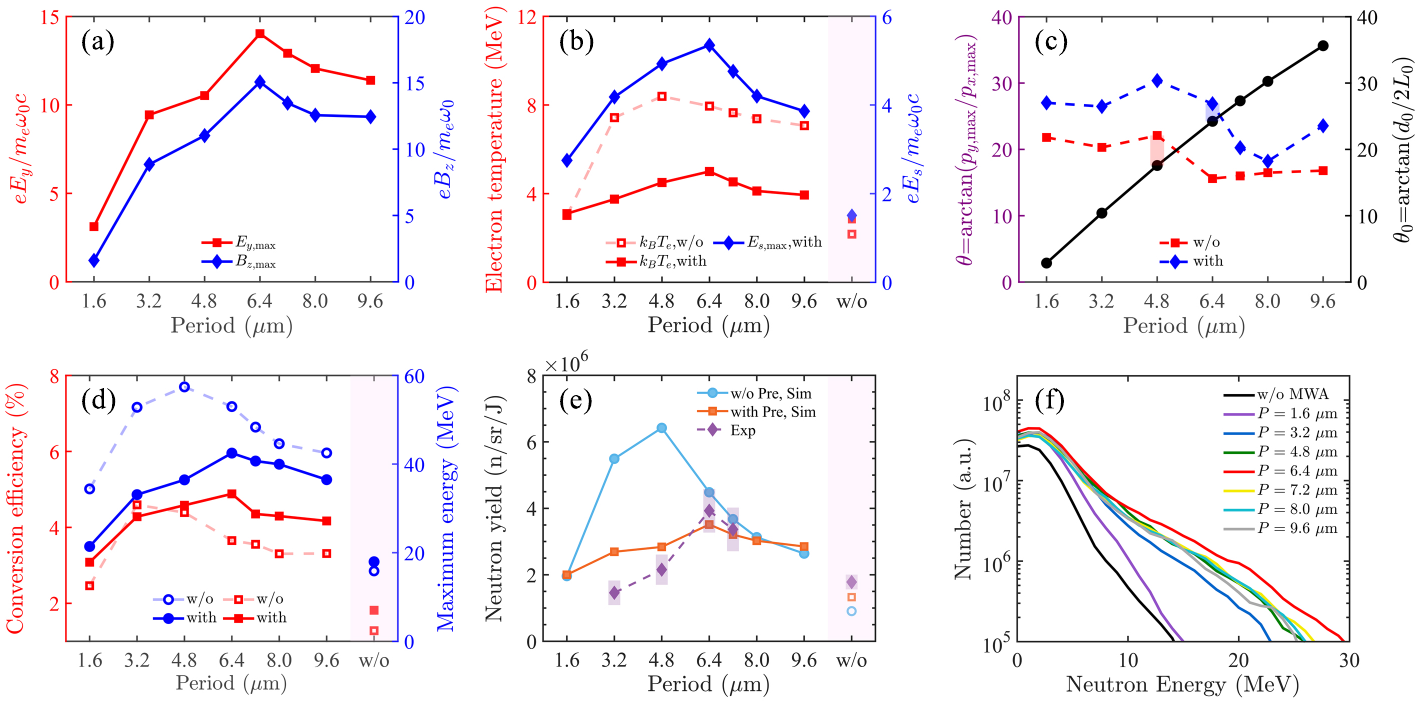} 
		\caption{Dependence of (a) the normalized laser fields at $\Delta t=60 T_0$ in MWAs, (b) the highest electron temperature $k_B T_h$ and sheath electric field, (c) the focusing angle of electron bunches $\theta = {\rm arctan} (p_{x,\rm max}/p_{y,\rm max})$ and the optimum angle $\theta_0 = {\rm arctan}(d_0/2L_0)$, (d) the proton cutoff energy $\mathscr{E}_{p,\rm cutoff}$ and energy conversion efficiency from the laser to the forward protons $\eta_P$ for the cases with (with pre) and without (w/o pre) the prepulse on array period $P$. (e) Neutron yield per joule and (f) energy spectra along the 0° direction at different $P$. The simulation results without the microwire-array (w/o) are given in (b) and (d)-(f) (pink background) for comparison and the experimental data is also shown in (e) for reference.}
		\label{fig4}
	\end{figure*}

For the DLA in MWA, the electron temperature $k_B T_h$ is dependent on the magnitudes of laser fields within the channels, where $k_B$ is the Boltzmann constant. Figure ~\ref{fig4}(a) illustrates the normalized strengths of laser fields at $\Delta t = 60 T_0$ from PIC simulations as a function of $P$. We see that $eE_y/m_e\omega_0c = 14$ and $eB_z/m_e\omega_0 = 15$ are highest at $P=6.4\ \mu \rm m$ and increased by $\sim$1.5 times with respect to the incident laser due to the intensification effect of local electric fields ~\cite{ji2016towards}. As expected, $k_B T_h$ is peaked at $P=6.4\ \mu \rm m$ and reaches about 4.8 MeV [see Fig.~\ref{fig4}(b)], which is beyond the ponderomotive temperature $k_B T_{h,\rm pond}=(\sqrt{1+a^{2}_{0}/2}-1)m_e c^2=3\ \rm MeV$ ~\cite{wilks2001energetic} and increases by more than twice compared to 1.8 MeV in the case of plane target, where $a_0 = 9.67$ is the normalized laser electric field. Since the sheath electric field $E_s$ scales as $\sqrt{T_h}$, $E_s$ exhibits similar distribution to $T_h$ with $P$ and is still strongest at $P=6.4\ \mu \rm m$ [also shown in Fig.~\ref{fig4}(b)]. It should be mentioned that, to achieve efficient proton acceleration, $L_0$ should be long enough to ensure sufficient acceleration distance. In addition, the angle $\theta = {\rm arctan} (p_{x,\rm max}/2p_{y,\rm max})$ [shown in Fig. ~\ref{fig3}(e)] should match with the optimum angle $\theta_0 = {\rm arctan}[L_0/(P-r_0)]$ in Fig.~\ref{fig1}(a), where $p_{x,\rm max}$ and $p_{y,\rm max}$ are the maximum longitudinal and transverse momenta of electrons~\cite{zou2017laser}. In this situation, the electron bunches can be effectively focused at the central region of the substrate (also seen in 3D simulation as shown in Appendix B), inducing a wider acceleration field. Figure~\ref{fig4}(c) shows the dependence of $\theta$ and $\theta_0$ on $P$ from PIC simulations for the MWA targets. One can see that $\theta$ is comparable to $\theta_0$ at $P=6.4\ \mu \rm m$ [the focusing effect of electron bunches can be easily observed in Fig.~\ref{fig3}(e)]. This indicates that  $P=6.4\ \mu \rm m$ should be the optimal MWA period at given $L_0=6.4\ \mu \rm m$ and $CR=10^{10}$ for proton acceleration. Figure~\ref{fig4}(d) shows the proton cutoff energy $\mathscr{E}_{p,\rm cutoff}$ and energy conversion efficiency $\eta_P$ from the laser to the forward protons at different $P$. We see that both $\mathscr{E}_{p,\rm cutoff}$  and $\eta_P$ are indeed peaked at $P=6.4\ \mu \rm m$, agreeing well with Figs.~\ref{fig2}(a) and \ref{fig2}(b). The highest $\mathscr{E}_{p,\rm cutoff}$ and $\eta_p$ reach about 42.5 MeV and $4.9\%$ respectively, significantly higher than the results obtained when $P\leq3.2\ \mu \rm m$. Moreover, compared with the plane target, the improvement is even twice as much. Figures~\ref{fig4}(e) and \ref{fig4}(f) show the neutron spectrum and yield per joule for the LiF converter with different $P$ from MC simulations. It is found that $\varepsilon_{n,\rm cutoff}$ and $N_n/\varepsilon_L$ are highest at $P=6.4\ \mu \rm m$ (i.e., the optimal period $P^{\rm Sim}_{\rm opt}=6.4\ \mu \rm m$), in good accordance with $P^{\rm Exp}_{\rm opt}$.

The ablation of the microwire structures by laser prepulses plays a significant role in electron heating and focusing, proton acceleration and neutron production for the MWA target. Figures~\ref{fig4}(b)-\ref{fig4}(e) also show that in the realistic environment with prepulses (with pre), $P^{\rm Exp}_{\rm opt}=P^{\rm Sim}_{\rm opt}(=6.4\ \mu \rm m)$ is larger than the optimal period from the simulations without prepulses (w/o pre) 
$P^{\prime}_{\rm opt} (\approx 4.8\ \mu \rm m)$. To interpret this discrepancy, we introduce an effective wire distance $d_{\rm eff}$, which characterizes the spacing between the critical plasma density positions of two adjacent expanded microwires, as presented in Fig.~\ref{fig3}(b). One can see that at $P=6.4\ \mu \rm m$, $d_{\rm eff}\approx 4.5\ \mu \rm m$, which is comparable to the actual distance of $d_0=P-r_0=3.8\ \mu \rm m$ at $P^{\prime}_{\rm opt} (\approx 4.8\ \mu \rm m)$. This indicates that $P$ should be increased accordingly due to the presence of laser prepulse. In addition, we find that $\mathscr{E}_{p,\rm cutoff}$, $\eta_p$ and $Y_n/\mathscr{E}_L$ all decrease owing to more laser energy loss in the preplasmas so high laser contrast is desirable. Note that the decrease of $k_B T_h$, $\mathscr{E}_{p,\rm cutoff}$, $\eta_p$ and $Y_n/\mathscr{E}_L$ with prepulses are exactly the opposite of the results from the plane target, where the preplasmas with suitable scale lengths may be beneficial for enhancing electron heating, proton acceleration and neutron generation~\cite{davis2008influence}.

\begin{figure}
		\centering
		\includegraphics[width=1\hsize]{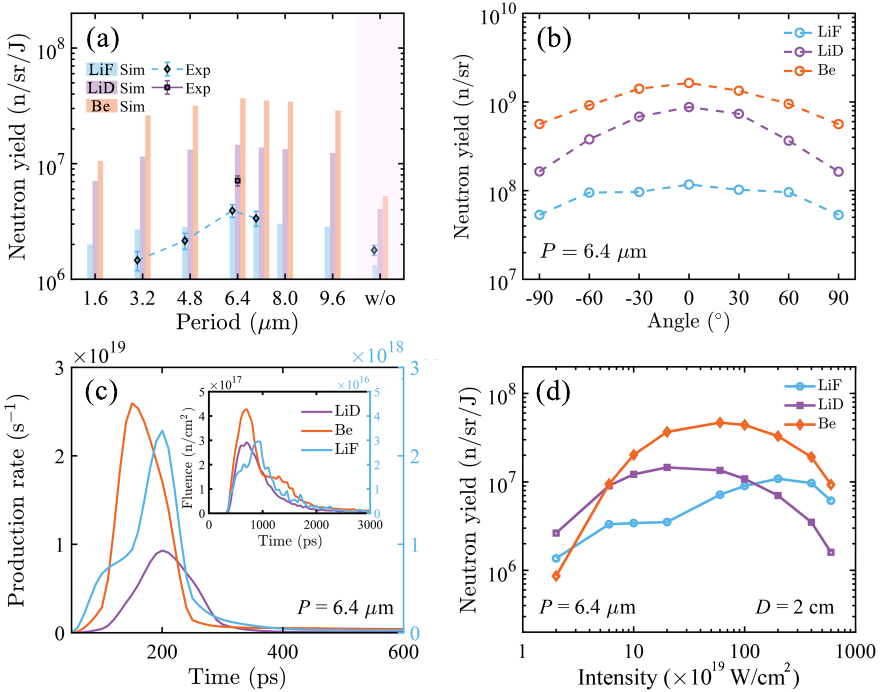} 
		\caption{Simulation results of three different material converters: LiF, LiD and Be. (a) Neutron yield per joule at different $P$. (b)-(d) Angular distribution of neutrons, temporal evolution of neutron production rate and neutron yield per joule of different laser intensities at $P=6.4\ \mu \rm m$. Here, the thickness of the converters is appropriately adjusted to 2~cm, while all other parameters remain unchanged.}
		\label{fig5}
	\end{figure}
    
Laser alignment is also critical for the laser interaction with MWA targets. Unlike the aforementioned PIC simulation where the laser strikes at the wire tip, the electron bunches are distributed symmetrically if the laser is incident along the central axis of the channel ($y=\pm P$) and and their focusing effect will be more distinct. Consequently, $\mathscr{E}_{p,\rm cutoff}$ is increased to $\sim$50 MeV at $P=6.4\ \mu \rm m$. However, $\eta_p$ has slightly decreased to $4.5\%$ because the wire number covered by the laser focal spot is reduced. As a whole, $Y_n/\mathscr{E}_L$ varies in some degree but still in a very delicate range (see Appendix C). For this axial incidence situation, the effect of the MWA will gradually weakens as $P$ increases. Especially when $P\gg2\sigma_0$, where $\sigma_0$ is the laser focal spot radius, the MWA target can even be regarded as a planar target. It is therefore necessary for $P$ to match the size of laser focal spot to achieve efficient laser proton acceleration and high neutron yield.
    
The converter is a key factor determining the quality of laser-driven neutron source. Figure~\ref{fig5}(a) shows the neutron yield per joule for three common converter material types (LiF, LiD and Be) with different $P$. We observe that when using LiD and Be converters, $P^{\rm Sim}_{\rm opt}$ corresponding to the highest $Y_n/\mathscr{E}_L$ remains at $6.4\ \mu \rm m$. In contrast, Be material is more efficient for neutron generation due to the high nuclear reaction cross-sections over a broader range of proton energies. $Y_n/\mathscr{E}_L$ can even reach 3.67~$\times$~$10^{7}$~n/sr/J [orange star in Fig.~\ref{fig2}(f)], which, to our knowledge, is the highest for ultrashort laser pulses at moderate intensities. Note that by using LiD converter, the forward directionality of neutrons could be significantly improved [see Fig.~\ref{fig4}(b)]. The angular distributions of neutrons for LiD and LiF converters were roughly consistent with the experimental observations. The temporal evolution of the neutron production rate obtained from three converters above is shown in Fig.~\ref{fig5}(b). One can see that the FWHM (full-width at half-maximum) pulse duration of neutron sources $\tau_n$ is as short as 70-100~ps at the moment of production. However, it will experience the time-of-flight broadening when these neutrons pass through the converter. On the rear surface of the converters, the measured $\tau_n$ has been extended to 460-670~ps. For the Be converter, the maximum neutron fluence can reach $4.2\times 10^{5}\ \rm n/cm^2$ [see the inset in Fig.~\ref{fig5}(c)] and the neutron peak flux is as high as $\sim9.1$~$\times$~$10^{18}\ \rm n/m^2/s$. Compared to that of the spallation neutron source ~\cite{taylor2007route,vogel2012brief}, this flux has increased by nearly two orders of magnitude.

To show the robustness of DLA scheme in neutron production, we further carry out a series of simulations with laser intensity ranging from $2\times 10^{19}\ \rm W/cm^2$ to $6\times 10^{21}\ \rm W/cm^2$ and $Y_n/\mathscr{E}_L$ from three converters above are shown in Fig.~\ref{fig5}(d). Considering that at higher laser intensities, the penetration depth of proton beam is longer, the thickness of the converters is thus adjusted to $D$ = 2 cm. We see that with the Be converter, $Y_n/\mathscr{E}_L$ is generally higher than $10^7$ n/sr/J at $I\geq6\times 10^{19}\ \rm W/cm^2$. When the laser intensity exceeds $1.3 \times 10^{21}\ \rm W/cm^2$, using a LiF converter can achieve a higher $Y_n/\mathscr{E}_L$ compared to the LiD converter because the stopping power of protons in the LiD material is relatively lower. We also evaluated the neutron generation at extreme output power of 3 PW in SILEX-II laser facility, and the thickness of Be converter was increased to 5 cm. $Y_{n,\rm max}/\mathscr{E}_L$ is even as high as $5.44 \times 10^7\ \rm n/sr/J$ [black star in Fig.~\ref{fig2}(f)], which has already approached the highest level that can be achieved using ps-class lasers with several hundred joules. It should be noted that similar structure, along with the Be converter, has also been explored in higher-power SULF facility~\cite{Dai2026}. 

In summary, we have experimentally demonstrated that for ultrashort laser pulses at moderate intensities, high neutron yield per joule exceeding (8.33~$\pm$~0.84)~$\times$~$10^{6}$~n/sr/J can be achieved by using 3D-printed microwire-array target. Self-consistent integrated simulations indicate that direct laser acceleration is highly efficient for improving proton acceleration and neutron generation. The optimal array period should be appropriately increased compared to the theoretical prediction due to the presence of the laser prepulse. A forward pulsed neutron source with yield up to 3.67~$\times$~$10^{7}~$n/sr/J can be obtained when the Be converter is employed under identical laser conditions. Our results can facilitate the application of laser-driven neutron sources in scenarios requiring high fluence and repetition rate, e.g., neutron resonance radiography and fusion material neutronic research.  

This work is supported by the National Natural Science Foundation of China (Grant Nos. 12275356, U22411281, 12135009 and 12375244) and the Natural Science Foundation of Hunan Province (2026JJ20011 and 2025JJ30002).

\nocite{*}


\newpage
\onecolumngrid
\begin{center}
\textbf{\Large End Matter}
\end{center}
\twocolumngrid

\textit{Appendix A---Details of self-consistent integrated simulation}: Self-consistent simulations are carried out by employing the radiation hydrodynamic (RHD) code Flash \cite{fryxell2000flash}, particle-in-cell (PIC) code Epoch \cite{arber2015contemporary} and Monte Carlo (MC) toolkit Geant4 \cite{shikhaliev2005quantum}. Here, we provide a detailed account of the parameter settings we used in the simulation process.

A. RHD Simulations: The Flash simulation is carried out in a two-dimensional (2D) Cartesian coordinate system, and the laser deposition is achieved through the ray-tracing projection of the Cartesian coordinate system \cite{fryxell2000flash}. The Courant-Friedrichs-Lewy (CFL) condition is set to 0.4, with the equation of state (EOS) being  from FEOS \cite{FAIK2018117} and the multi-group opacity tables from the SNOP non-LTE atomic physics model \cite{FAIK2018117}. The multi-group radiation transport model consist of a total of 20 energy groups with their energy range covering from 1 eV to 10 keV. This ensures that the radiative cooling and energy transport processes can be accurately described in the simulation of the preplasma expansion and heating. Flash code uses an adaptive mesh refinement (AMR) scheme and the coarsest/finest mesh size used is 0.09375/0.01172 $\mu \rm m$ in the axial (x) and radial (y) directions. For the parameter setting of the incident laser, we refer to the prepulse setting of the SILEX-II facility \cite{hong2021commissioning}. The laser contrast ratio is $10^{10}$ in $t$ = 0-100 ps, and the intensity gradually rise until $t =120$ ps. The 3D-printed MWA parameters are inputted into FLASH. The MWA consists of a polymer material of mass density 1.17 $\rm g/cm^3$, with an initial temperature of 290 K. Outflow boundary conditions are used along both the x and y directions for the simulations.

B. PIC Simulations: The density profile of MWAs obtained from Flash simulation is applied as the initial conditions for 2D PIC simulations with Epoch \cite{arber2015contemporary}. The MWAs are regarded as being fully ionized due to the ablation of laser prepulse. In the Epoch simulation, the simulation box is $x \times y = 100\ \mu \rm m \times 30\ \mu \rm m$ with $10000 \times 1500$ cells and 36 macro particles per cell. The target consists of a MWA, a SiN flat substrate and a thin CH contaminant layer. The targets are fully ionized as cold plasma, which is initially neutral. The SiN substrate has the ion densities of ${\rm Si^{14+}:N^{14+}}=12n_c$, and is of thickness 200 nm. The CH contaminant layer with a thickness of 24 nm and the electron density of $n_e=40n_c$ \cite{qin2022high}. The MWA target of the electron density $n_e=210n_c$ is located between $x=14\lambda_0$ and $x=20\lambda_0$. The main pulse, of wavelength $\lambda_0=0.8\ \mu\rm m$ and period $T_0=2.67\ \rm fs$, is focused on the front surface of the SiN substrate at an incidence angle of $5^\circ$. Its temporal and spatial waveform follows a Gaussian distribution, with a pulse duration of 30 fs and a focal spot radius of $5\ \mu \rm m$. The intensity of the main pulse is $2\times10^{20}\ \rm W/cm^2$, and the normalized amplitude corresponds to $a_0=eE_0/m_e\omega_0c=9.67$, respectively.

\begin{figure}[!b]
    \centering
    \includegraphics[width=1\linewidth]{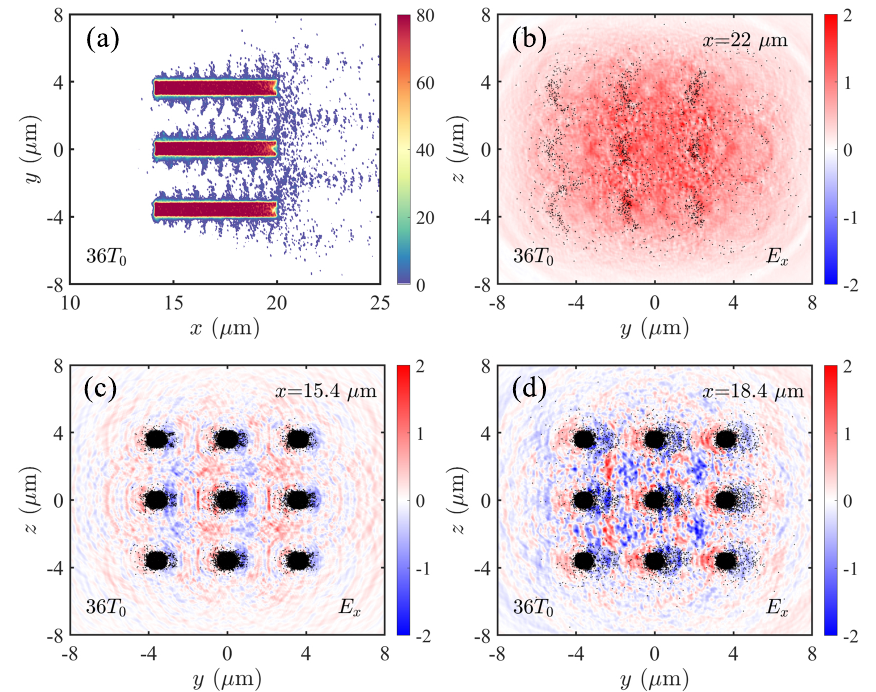}
    \caption{3D PIC simulation results at $\Delta t=36T_0$. Distribution of (a) the electron density along the $ z = 0$ cross-section, the sheath electric field along the $x = 22\ \mu \rm m$ cross-section as well as the longitudinal electric field along $x = 15.4\ \mu \rm m$ and $x = 18.4\ \mu \rm m$ cross-sections, respectively.}
    \label{figS1}
\end{figure}

C. MC Simulations: All the positions, momenta and energies as well as particle weight information of the ions from Epoch simulations are once again imported into the 3D Geant post-processing program within the MC framework. The $\rm QGSP\_BIC\_ALLHP$ physics list was employed, which is created using high-precision charged particle physics models and is suitable for cases with energies below 200 MeV, enabling accurate description of nuclear reaction processes between protons and target materials such as LiF, LiD, and Be. The nuclear reaction cross section data are derived from the TENDL library. The laser-accelerated ion beams can be regarded as a point source since its spatial size is much smaller than that of the converter. To ensure the number of injected particles in the 2D simulation matches that in the 3D scenario, we multiply the particle number obtained from the 2D simulation by a geometric conversion factor \cite{qin2022high} $ N_{3D}/N_{2D} = (L\times L) / (L\times h) = L/h = 3.0 \times 10^{-5}$. Here, $L= 30\ \mu \rm m$ is the y-direction size of the target in the 2D PIC simulation (we assume its cross section in 3D space is a square of area $L \times L$ ), and $h = 1\ \mu \rm m$ is the default unit thickness in the third dimension (the $z$-direction). The temporal distribution of neutron generation were recorded to analyze the neutron duration. Two distinct definitions of duration were evaluated: (1) the distribution of the neutron birth time, corresponding to the exact moments of neutron production through nuclear reactions; and (2) the distribution of time at which neutrons exit the rear surface of the converter, representing a more experimentally relevant measure that approximates the temporal signal detectable by a detector positioned behind the converter. These two definitions provide complementary insights into the temporal behavior of neutron production: the former directly reflects the dynamics of nuclear reactions induced by the interaction of the ion beam with the converter material, while the latter accounts for time broadening effects arising from neutron transport processes within the converter medium. To investigate the influence of laser prepulse on neutron production, we also conducted the full-process simulations without laser prepulses. The role of converter material is then discussed, and three common material LiF, LiD and Be are chosen for comparison. In the above simulations, we vary one of the parameters in turn, and other parameters remain unchanged.

\textit{Appendix B---3D PIC simulation of laser interacting with MWA target}: To demonstrate that the focusing effect of electron bunches also exists in actual 3D space, we also conducted 3D PIC simulations. For simplicity, we only consider the case without laser prepulse, and the MWA period is chosen to $P = 3.6\ \mu \rm m$ according to Fig. 4(d). In the Epoch simulation, the simulation box is $x \times y \times z = 35{\ \mu \rm m} \times 24{\ \mu \rm m} \times 24{\ \mu \rm m}$ with $1050 \times 360 \times 360$ cells and 9 macro particles per cell. To save the computing resources, the electron density of MWA target is reduced to $n_e=84n_c$.

Figure \ref{figS1} shows the 3D PIC simulation results at $\Delta t=36T_0$. We observe that in 3D geometry, the electron bunches are still clearly concentrated in front of the substrate, and intense sheath electric field on the rear surface can induced as they pass through the substrate. In addition, we note that the longitudinal electric fields also appeared inside the channels, which originate from the electric field component of the incident laser and charged separation fields from the pulled out electrons. They distributed around the microwire surfaces at an early time, and gradually occupied the central region of the channels accompanying with the focusing electron bunches. The MWA, in nature, is different from multi-channel (the length along the $z$ direction is infinite) or microtube structures which can be regarded as rectangular or cylindrical plasma waveguides. In the plasma waveguides, the longitudinal electric fields are mainly from the high-order transverse magnetic modes excited inside the channels, which are more helpful for the charged particle acceleration~\cite{zou2017laser,yi2016,zou2019}.

\begin{figure}[!t]
    \centering
    \includegraphics[width=1\linewidth]{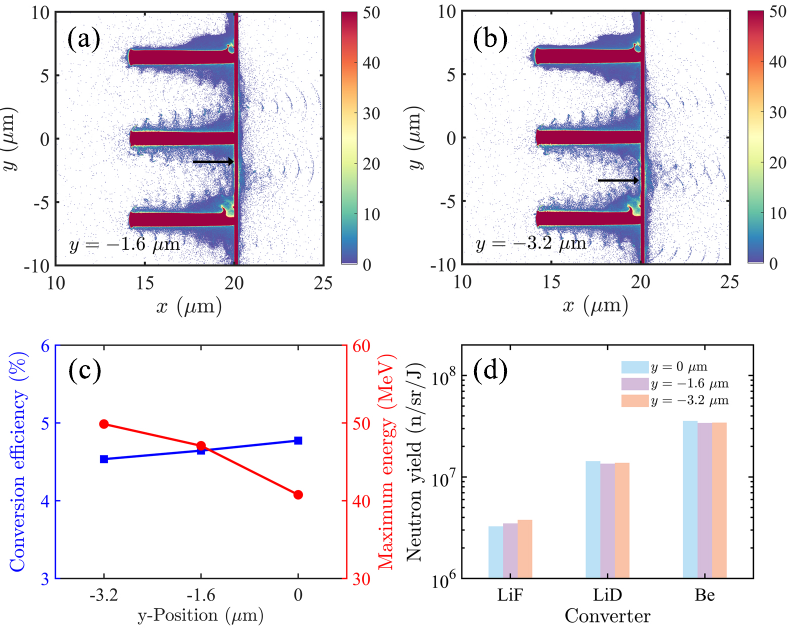}
    \caption{2D PIC simulation results. (a), (b) Distribution of the electron density, (c) energy conversion efficiency from the laser to the forward ($p_x>0$) protons and proton cutoff energy, and (d) neutron yield per joule from different converters when the laser is incident along the $y = -1.6\ \mu \rm m$ and $-3.2\ \mu \rm m$ directions, respectively.}
    \label{figS2}
\end{figure}

\textit{Appendix C---Effect of laser alignment on neutron generation}: We now discuss the effect of laser alignment on proton acceleration and neutron generation in detail. Different from the situation where the laser strikes at the microwire tip $(y = 0)$ at $P= 6.4\ \mu \rm m$, we consider two typical cases in which the laser is incident along the $y = -1.6\ \mu \rm m$ and $-3.2\ \mu \rm m$ directions, respectively. Other parameters are the same as the previous. The simulation results at $\Delta t=36T_0$ are shown in Fig. \ref{figS2}. One can see that in contrast the distribution of electron bunches are almost symmetrical when the laser is incident along the central axis of the channel. Their focusing effect becomes more distinct, leading to an increase of proton cutoff energy $\varepsilon_{p,\rm cutoff}$ from $\sim$40 MeV to $\sim$50 MeV. Conversely, the energy conversion efficiency from the laser to the forward protons $\eta_p$ is slightly decreased to $4.5\%$ since the effective wire number covered by the laser focal spot is reduced. As a result, the neutron yield per joule is still comparable although the laser loading point is different.

\end{document}